# Comparison of the number of history in Monte Carlo Simulation Programs

E. Uyar[a*], Z.A. Günekbay[a]

[a)]Gazi University, Faculty of Sciences, Department of Physics, Ankara, Turkey

[*]Corresponding author: Esra Uyar
E-mail: esrauyar@gazi.edu.tr
Tel: +90 312 202 12 36
Fax: +90 312 212 22 79

**ABSTRACT**

The use of the Monte Carlo technique in a reliable and inexpensive way without the need for a standard radioactive source in determining the detector efficiency is becoming widespread every passing day. It is important to model the detector with the real dimensions for an accurate and precise results for the method. Another parameter as important as detector modeling is the number of histories in the simulation code examined in this study. In this study, the effect of the number of histories on the efficiency was examined in detail using different simulation codes. The results obtained in this work, at least $10^7$ particle numbers should be used in all three programs where the uncertainty is below 1%. If the existing facilities are sufficient, it can be increased to $10^8$s in case of having a more equipped and fast computer. However, going higher than this value does not make any sense as seen from the study.

## 1. Introduction

Monte Carlo (MC) simulation is a statistical technique for directly simulating a physical process. The basis of the MC method is a random number generator consisting of random numbers in the range (0, 1). Since such numbers are generated by deterministic algorithms, they are untruly random. However, such pseudo-random numbers are statistically indistinguishable from real random numbers that are evenly distributed in the range (0, 1) and are independent of each other [1]. Simulating particle transport in three-dimensional MC codes are widely used in various research and development fields such as radiotherapy, radiation shielding, medical physics, nuclear technology, accelerator design, astrophysics applications. detector modeling [2]. The use of the MC method in gamma-ray spectrometry, where HPGe detectors are used, is becoming more common day by day. Accurate modeling of the detection chain via the MC method is crucial for obtaining quality data from detectors and for the design of experiments. MC modeling of radiation detectors is a widely accepted numerical method for detector characterization. For example, MC programs are a excellent guide for the characterization of the detectors by determining the dead layer thickness, which is a time-varying parameter in HPGe detectors [3]. In gamma-ray spectrometric studies, the MC method is mostly used to obtain the efficiency value of the detector [4–7].

Since this method is a statistical process in which random numbers are used, keeping the number of histories as high as possible allows us to obtain more meaningful results. But there is no exact value regarding the number of histories. For example, Ordonez et al. set the number of histories in each simulation as 20 million to achieve statistical errors of less than 1.5% [6]. Azli et al. used 100 million particles to achieve a relative error of less than 1% in the calculated efficiency [8]. Subercaze et al. 3 million [9]; Miroslav et al. used 1 million particles in their study [10]. Therefore, as can be seen from the literature, a number of histories ranging from 1 million to 100 million were used. Here, the performance and features of the computer used in the calculation are very important.

In this study, the effect of history number on efficiency calculation with the MC method was investigated using three different MC programs. For this purpose, the efficiency values of $10^5$, $10^6$, $10^7$ and $10^8$ particle numbers were calculated and the effect of the particle number on the efficiency was investigated for the 59.5, 383.9, 661.7, 1173.2 and 1332.5 keV peaks at both low and high energy.

## 2. Materials and Methods

MC codes used in gamma-ray spectrometry fall into two categories: specialized codes in gamma-ray spectrometry, mostly written specifically for efficiency calculations, and multi-purpose MC codes for a wide variety of applications. The PHITS used in this study is a multi-purpose code for all kinds of applications; GESPECOR and DETEFF are specialized purpose MC codes.

## 2.1 PHITS Monte Carlo simulation code

PHITS is a multi-purpose MC simulation code for particle transport that was created in collaboration between JAEA, RIST, KEK, and numerous other institutions. With the use of several nuclear reaction models and nuclear data libraries, it can be interested in the transport of all particles over various energies [11]. The parameters for the history number in PHITS are maxcas and maxbch (Fig.1). The total number of histories is equal to the product of maxcas, the number of particles per batch, and maxbch, the number of batches. It is recommended to set the maxbch value greater than or equal to 10 to obtain reliable results. A larger maxbch provides more reliable statistical uncertainties, but may require a longer computation time.

```
[ Parameters ]
 icntl    =          0    # (D=0) 3:ECH 5:NOR 6:SRC 7,8:GSH 11:DSH 12:DUMP
 maxcas   =    1000000    # (D=10) number of particles per one batch
 maxbch   =         10    # (D=10) number of batches
```

Fig.1. Defining the number of histories in the PHITS MC code

## 2.2. GESPECOR

GESPECOR (GErmanium SPEctra CORrection) is a special-purpose MC-based software developed for calculating true coincidence summing and self-absorption effects, especially full energy peak efficiency in gamma-ray spectrometry. In GESPECOR, the number of histories is determined by the number of runs entered in the window that opens automatically in the system before starting the simulation (Fig. 2).

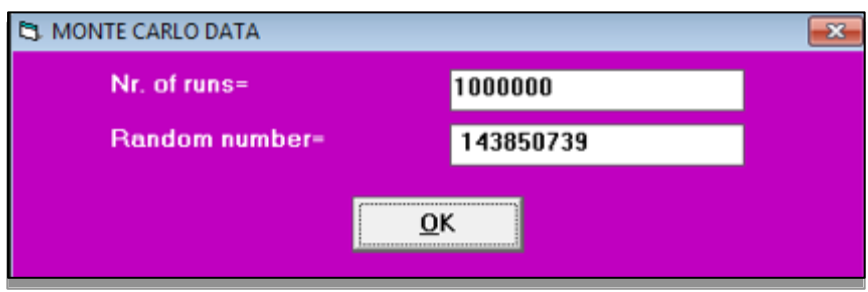

Fig.2. Defining the number of histories in the GESPECOR MC code

## 2.3. DETEFF

DETEFF; it is a user friendly MC program for calculating the full energy peak efficiency in gamma-ray detectors such as NaI, CsI, Ge(Li), HPGe and Si(Li) [12]. In DETEFF, the parameters for the number of histories are experiments and number of photons in the Statistics tab (Fig.3). The total number of histories is equal to the value in the experiments multiplied by the number of photons.

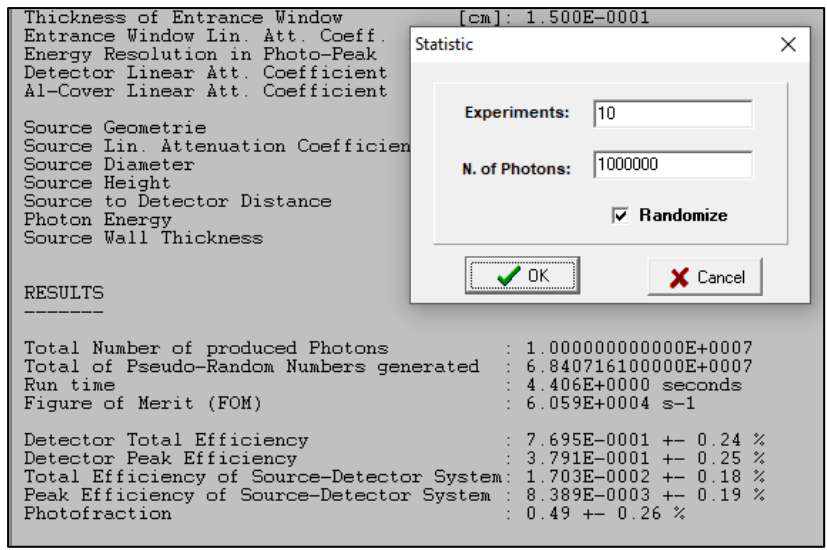

Fig.3. Defining the number of histories in the DETEFF MC code

## 3. Results and Discussion

Measurements were taken at $10^5$, $10^6$, $10^7$, $10^8$ using PHITS, GESPECOR and DETEFF MC codes and $^{241}$Am (59.54 keV), $^{133}$Ba (383.85 keV), $^{137}$Cs (661.66 keV) and $^{60}$Co (1173.23 keV and 1332.49 keV) peaks. The efficiency values taken at different particle numbers, with their uncertainties, are given in Table 1.

**Table 1.** Efficiency values obtained with PHITS, GESPECOR and DETEFF for different history numbers

|  | Nuclide | Energy (keV) | PHITS (Uncertainty %) | GESPECOR (Uncertainty %) | DETEFF (Uncertainty %) |
|---|---|---|---|---|---|
| $10^5$ | $^{241}$Am | 59.54 | 0.00455 (0.047) | 0.00539 (0.057) | 0.00529 (0.840) |
|  | $^{133}$Ba | 383.85 | 0.01095 (0.030) | 0.01086 (0.270) | 0.01150 (1.770) |
|  | $^{137}$Cs | 661.66 | 0.00766 (0.036) | 0.00803 (0.360) | 0.00855 (2.440) |
|  | $^{60}$Co | 1173.23 | 0.00568 (0.042) | 0.00592 (0.510) | 0.00628 (3.780) |
|  | $^{60}$Co | 1332.49 | 0.00517 (0.044) | 0.00552 (0.370) | 0.00611 (3.410) |
| $10^6$ | $^{241}$Am | 59.54 | 0.00475 (0.015) | 0.00539 (0.012) | 0.00534 (0.440) |
|  | $^{133}$Ba | 383.85 | 0.01082 (0.010) | 0.01089 (0.110) | 0.01115 (0.830) |
|  | $^{137}$Cs | 661.66 | 0.00796 (0.011) | 0.00807 (0.160) | 0.00848 (0.420) |
|  | $^{60}$Co | 1173.23 | 0.00589 (0.013) | 0.00594 (0.140) | 0.00616 (1.000) |
|  | $^{60}$Co | 1332.49 | 0.00543 (0.014) | 0.00550 (0.120) | 0.00577 (0.910) |
| $10^7$ | $^{241}$Am | 59.54 | 0.00483 (0.005) | 0.00539 (0.006) | 0.00530 (0.130) |
|  | $^{133}$Ba | 383.85 | 0.01090 (0.003) | 0.01089 (0.024) | 0.01126 (0.270) |
|  | $^{137}$Cs | 661.66 | 0.00815 (0.004) | 0.00807 (0.042) | 0.00842 (0.210) |
|  | $^{60}$Co | 1173.23 | 0.00599 (0.004) | 0.00593 (0.068) | 0.00615 (0.250) |
|  | $^{60}$Co | 1332.49 | 0.00557 (0.006) | 0.00551 (0.042) | 0.00571 (0.150) |
| $10^8$ | $^{241}$Am | 59.54 | 0.00483 (0.001) | 0.00539 (0.004) | 0.00530 (0.050) |
|  | $^{133}$Ba | 383.85 | 0.01091 (0.001) | 0.01088 (0.012) | 0.01125 (0.070) |
|  | $^{137}$Cs | 661.66 | 0.00815 (0.001) | 0.00807 (0.011) | 0.00838 (0.060) |
|  | $^{60}$Co | 1173.23 | 0.00600 (0.001) | 0.00592 (0.015) | 0.00616 (0.100) |
|  | $^{60}$Co | 1332.49 | 0.00559 (0.001) | 0.00551 (0.011) | 0.00571 (0.100) |

As shown in the Table 1, the increase in the number of history does not cause a linear change in the efficiency. While the increase in the number of histories decreases the efficiency in some energies, the efficiency increases in some energies. However, the percentage uncertainty values decrease as the number of history increases in all MC codes, that is, they improve. According to Table 2, where the relationship between the history numbers is examined, the percentage difference decreases as the number of particles increases in all MC codes.

**Table 2.** Percentage difference values from PHITS, GESPECOR and DETEFF

|     | Nuclide | Energy (keV) | PHITS* | GESPECOR* | DETEFF* |
|-----|---------|--------------|--------|-----------|---------|
| n:5 | $^{241}$Am | 59.54 | 4.37 | 0.02 | 0.98 |
|     | $^{133}$Ba | 383.85 | 1.21 | 0.25 | 3.04 |
|     | $^{137}$Cs | 661.66 | 3.90 | 0.53 | 0.78 |
|     | $^{60}$Co | 1173.23 | 3.84 | 0.30 | 1.85 |
|     | $^{60}$Co | 1332.49 | 5.11 | 0.36 | 5.57 |
| n:6 | $^{241}$Am | 59.54 | 1.75 | 0.01 | 0.64 |
|     | $^{133}$Ba | 383.85 | 0.79 | 0.04 | 0.99 |
|     | $^{137}$Cs | 661.66 | 2.34 | 0.05 | 0.70 |
|     | $^{60}$Co | 1173.23 | 1.50 | 0.25 | 0.19 |
|     | $^{60}$Co | 1332.49 | 2.59 | 0.15 | 1.04 |
| n:7 | $^{241}$Am | 59.54 | 0.05 | 0.01 | 0.08 |
|     | $^{133}$Ba | 383.85 | 0.25 | 0.06 | 0.09 |
|     | $^{137}$Cs | 661.66 | 0.25 | 0.04 | 0.32 |
|     | $^{60}$Co | 1173.23 | 0.37 | 0.04 | 0.16 |
|     | $^{60}$Co | 1332.49 | 0.42 | 0.03 | 0.14 |

*: The percent difference values between the particle numbers were calculated according to the $\frac{10^n - 10^{n+1}}{10^n} \times 100$ equation.

The biggest difference in the PHITS MC code occurred at n:5, that is, between $10^5$ and $10^6$ history numbers (up to 4.4%). When $10^6$ and $10^7$ data are examined, it is seen that the difference values are smaller and close to each other (Fig. 4).

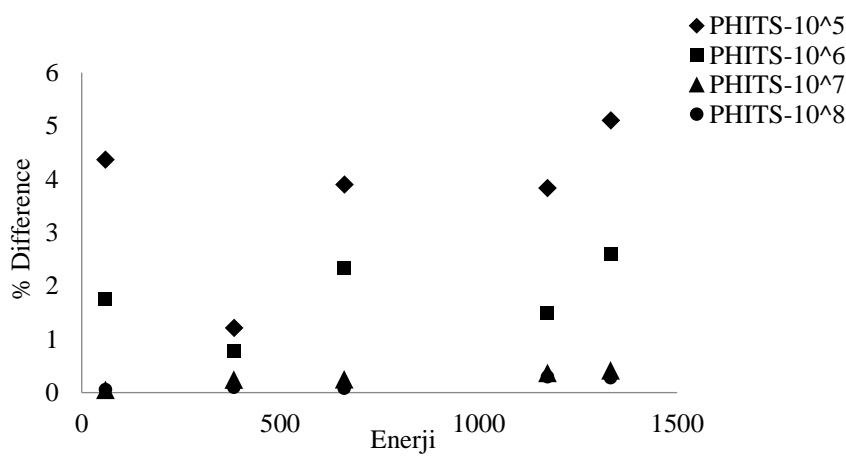

Fig.4. Variation of percent difference values between particle numbers obtained with PHITS according to energy

When the percent difference values between the particle numbers calculated according to the Equation (1) were examined for GESPECOR, an uncertainty of less than 1% was obtained in all particle numbers. The lowest uncertainty was obtained between $10^7$ and $10^8$, as expected (Fig. 5).

$$\frac{10^n - 10^{n+1}}{10^n} \times 100 \qquad (1)$$

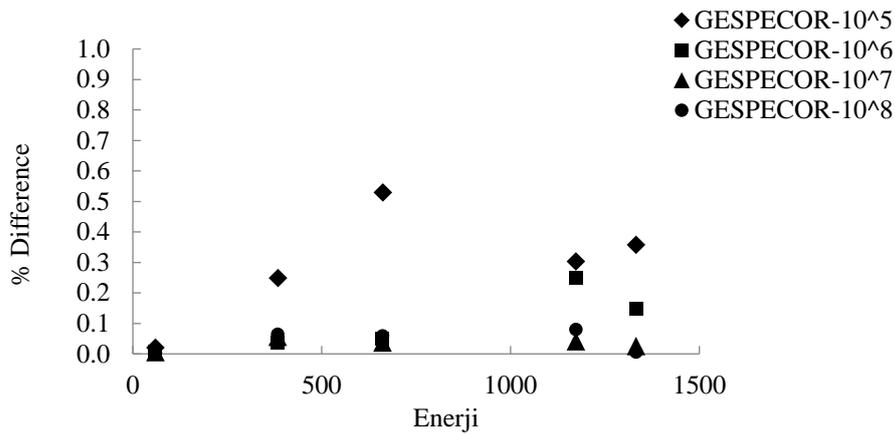

Fig.5. Variation of percent difference values between particle numbers obtained with GESPECOR according to energy

The biggest difference in the DETEFF MC code occurred at n:5, that is, between the historical numbers 105 and 106 (up to 5.6%). When the $10^7$ and $10^8$ data are examined, it is seen that the difference values are less than 1% (Fig. 6).

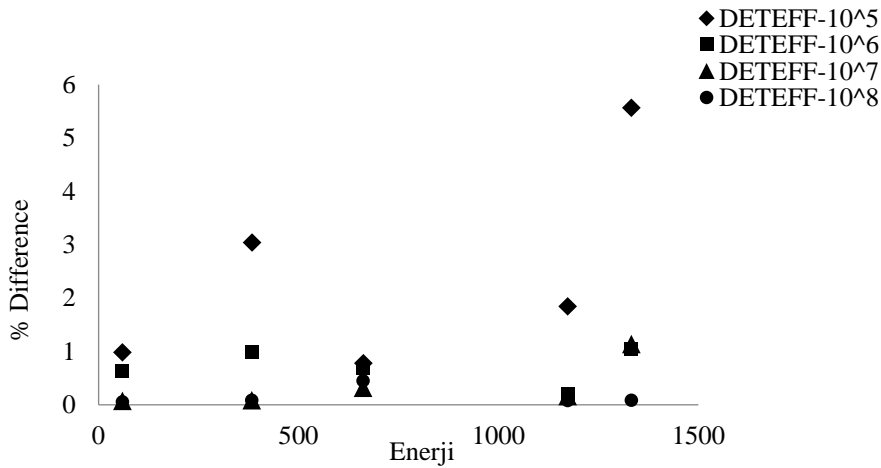

Fig.6. Variation of percent difference values between particle numbers obtained with DETEFF according to energy

## 4. Conclusions

In this study, the effect of the number of histories on the efficiency was investigated with different MC codes. Efficiency values were obtained at energies in the range of 59.5-1332.5 keV by using $10^5$, $10^6$, $10^7$ and $10^8$ particle numbers. Efficiency values were obtained for each code at the same particle count and the same energy at varying times ranging from a few seconds to several hours. Therefore, it is seen that the execution times of each code is different from each other. It has been observed that the dedicated packages GESPECOR and DETEFF give much faster results on average than general-purpose MC code PHITS. This is because dedicated codes use various variance reduction techniques, only transporting photons and electrons. In DETEFF and GESPECOR, the percentage difference between the particle numbers is less, but the repeatability is low. In other words, different values are obtained because of repeating the same simulation. General-purpose packages like PHITS are at a disadvantage in this respect due to their more complex physics and particle tracking [13]. In this study, it was determined that at least $10^7$ particle numbers should be adjusted to obtain good statistics in simulations where gamma-ray spectrometric calculations are made.